# Self-regulated radius of spontaneously formed GaN nanowires in molecular beam epitaxy

Sergio Fernández-Garrido,[*],[†] Vladimir M. Kaganer,[†] Karl K. Sabelfeld,[†],[¶] Tobias Gotschke,[†] Javier Grandal,[‡],[§] Enrique Calleja,[‡] Lutz Geelhaar,[†] and Oliver Brandt[†]

*Paul-Drude-Institut für Festkörperelektronik, Hausvogteiplatz 5–7, 10117 Berlin, Germany, and ISOM and Dpt. de Ingeniería Electrónica, ETSI Telecomunicación Universidad Politécnica de Madrid, 28040 Madrid, Spain*

E-mail: garrido@pdi-berlin.de

**Abstract**

We investigate the axial and radial growth of GaN nanowires upon a variation of the Ga flux during molecular beam epitaxial growth. An increase in the Ga flux promotes radial growth without affecting the axial growth rate. In contrast, a decrease in the Ga flux reduces the axial growth rate without any change in the radius. These results are explained by a kinetic growth model that accounts for both the diffusion of Ga adatoms along the side facets towards the nanowire tip and the finite amount of active N available for the growth. The model explains the formation of a new equilibrium nanowire radius after increasing the Ga flux and provides an explanation for two well known but so far not understood experimental facts: the necessity

---
[*]To whom correspondence should be addressed
[†]Paul-Drude-Institut für Festkörperelektronik
[‡]ISOM and Dpt. de Ingeniería Electrónica, ETSI Telecomunicación Universidad Politécnica de Madrid
[¶]Permanent address: Institute of Computational Mathematics and Mathematical Geophysics, Russian Academy of Sciences, Lavrentiev Prosp. 6, 630090 Novosibirsk, Russia
[§]Present address: Paul-Drude-Institut für Festkörperelektronik, Hausvogteiplatz 5–7, 10117 Berlin, Germany

of effectively N-rich conditions for the spontaneous growth of GaN nanowires and the increase in nanowire radius with increasing III/V flux ratios.

The spontaneous growth of GaN and ZnO nanowires (NWs) is fundamentally different from the well-established vapor-liquid-solid growth mechanism where liquid or solid metal particles are used to collect the precursor species and induce uniaxial growth.[1] The latter method enables the growth of uncoalesced NWs with high aspect ratios, but suffers from inherent difficulties such as the spontaneous switching between different crystal phases during growth, and the principle inability to fabricate axial heterostructures with chemically abrupt interfaces.[2–6]

In contrast, GaN and ZnO do not require any metal particles inducing uniaxial growth but spontaneously form NWs under specific growth conditions.[7–11] In addition, they can be grown on a wide variety of substrates, including Si or glass, without detrimental effects for their optical and structural properties.[9,12–17] Consequently, they have attracted great interest for the fabrication of nanoscale devices such as light emitters, solar water splitting cells, chemical sensors, and photovoltaic cells on inexpensive and large area substrates.[9,18–24] However, the development of NW based GaN and ZnO nanoscale devices is hampered by the incomplete understanding of the physical mechanisms governing the spontaneous nucleation and growth of NWs, and the resulting lack of control of their morphological properties.

In plasma-assisted molecular beam epitaxy (PA-MBE), GaN NWs can be exclusively grown under effectively (taking Ga desorption into account) N-rich conditions (III/V<1) at higher temperatures (> 750 °C) than those commonly used for epitaxial GaN films.[8,25,26] Under these conditions, GaN NWs crystallize in the wurtzite structure with the polar (0001) axis parallel to the growth direction. In the absence of morphological or structural defects of the substrate, GaN NWs are irrevocably N-polar.[11,27–31] The final NW radius scales with the impinging III/V flux ratio.[26,32,33] Neither the necessity of effectively N-rich conditions for obtaining NW growth nor the dependence of their radius on the III/V ratio is understood.

In this letter, we present experiments designed to elucidate the origin of these phenomena, and thus to clarify the mechanisms governing the NW morphology. Since the impinging III/V flux ratio may influence not only the growth but also the nucleation of GaN NWs,[26,34] we utilize fully-developed NWs with a well-defined height and radius as a template for our systematic investigation of the changes in axial and radial growth induced by a change in the Ga flux during growth. The experimental results are analyzed by a kinetic growth model that accounts for both the diffusion of Ga adatoms towards the NW tip and the finite amount of active N available for the growth. This latter limitation has so far been ignored because the growth of GaN NWs is carried out under N excess,[35–41] but is found to be an essential ingredient for explaining our observations. When the actual III/V ratio is explicitly taken into account, the final NW radius becomes self-regulated and depends on the impinging III/V flux ratio. The present model also predicts that, in order to avoid the formation of a compact layer as result of NW coalescence, effectively N-rich growth conditions are required. Finally, we combine our growth model with the nucleation study published by Consonni *et al.* in Ref. 42 to provide a comprehensive description of the spontaneous formation of GaN NWs in PA-MBE.

All samples were grown in a DCA Instruments P600 MBE system equipped with two radio frequency $N_2$ plasma sources for active N and two solid-source effusion cells for Ga. The angle of the cells with respect to the substrate normal is 44°. Cross-sectional scanning electron microscopy (SEM) of thick GaN(0001) films grown under slightly N- and Ga-rich conditions at low temperatures (680°C) was used to calibrate the Ga and N fluxes $\Phi_{Ga}$ and $\Phi_N$ in GaN-equivalent growth rate units (nm/min).[43] A growth rate of 1 nm/min is equivalent to 0.064 ML/s, where 1 ML of Ga corresponds to an areal density of $1.14\times10^{15}$ adatoms /cm$^2$. The substrate temperature was measured by an optical pyrometer. The as-received Si(111) substrates were chemically cleaned with solvents and etched using diluted (5%) HF. Prior to growth, 40 MLs of Ga were deposited at 550°C and flashed-off at a temperature of 1050°C for 60 s to remove any residual $SiO_2$ from the surface. All samples were grown at 815°C with $\Phi_N$ = (10.8 ± 0.5) nm/min. The desorbing Ga flux during the experiments $\Phi_{Ga}^{QMS}$ was monitored in-situ by line-of-sight quadrupole mass spectrometry (QMS).[44] The QMS response was calibrated in GaN-equivalent growth rate units as described in Ref. 45. During growth, the samples were rotated at 5 rpm. The NW areal density, the average NW radius, and the surface area fraction covered by the GaN NWs were determined by a statistical analysis of plan-view SEM images using ImageJ.[46]

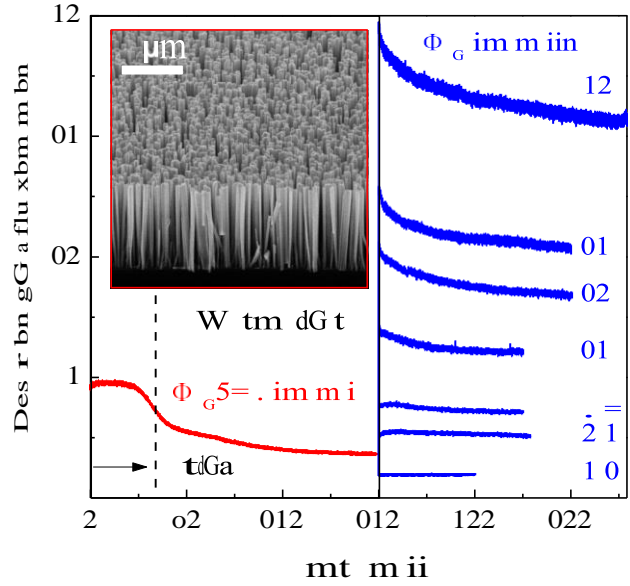

Figure 1: Time evolution of the desorbing Ga flux during the two-step growth experiments. The solid line on the left represents the desorbing Ga flux during the growth of the NW template with $\Phi_{Ga}$ = 4.8 nm/min. The vertical dashed line indicates the average delay time for the formation of GaN NWs. The solid lines on the right (after 180 min) show the desorbing Ga flux for different samples during the second step, where the NW template was overgrown using values of $\Phi_{Ga}$ between 2.1 and 26 nm/min. The inset shows a bird-eye SEM image of the NW template.

To evaluate the influence of the impinging Ga flux on the axial and radial growth rates of fully-developed GaN NWs, we perform a series of two-step experiments. The first step is identical for all samples and consists in the growth of a GaN NW template with $\Phi_{Ga}$ = (4.8 ± 0.6) nm/min. As shown in the inset of Figure Figure 1, the NW template consists of 1.5 $\mu$m high, well separated and homogeneous NWs with an areal density of $3.5 \times 10^9$ cm$^{-2}$ and an average radius of 43 nm. During the second step, the Ga flux $\Phi_{Ga}$ is changed and varied among the samples between (2.1 ± 0.3) and (26 ± 3) nm/min. In contrast to a one-step growth experiment,[26] this approach allows us to unambiguously elucidate the influence of $\Phi_{Ga}$ on the axial and radial growth rates during the growth of fully-developed NWs because the nucleation stage was identical for all samples.

Figure Figure 1 shows the time evolution of the desorption flux $\Phi_{Ga}^{QMS}$ during the two steps of the growth experiment. During the growth of the NW template, $\Phi_{Ga}^{QMS}$ exhibits the characteristic behavior reported in Refs. 47 and 48: an initial stage associated to the incubation time[34,35,42] where $\Phi_{Ga}^{QMS}$ is constant, a second stage where $\Phi_{Ga}^{QMS}$ rapidly decreases due to the formation of

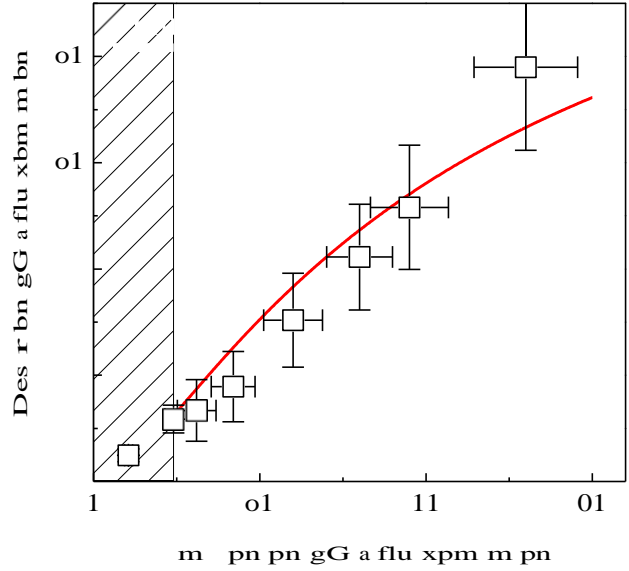

Figure 2: Steady-state desorbing Ga flux during the second growth stage as a function of the impinging Ga flux. The squares represent the experimental data and the solid line is the model calculation by Eq. Eq. (7). The hatched area corresponds to values of $\Phi_{Ga}$ lower than the one used for growing the NW template (4.8 nm/min). The experimental value for the NW template was also included.

GaN NWs, and a final stage related to the elongation of GaN NWs where $\Phi_{Ga}^{QMS}$ reaches steady-state conditions. According to the QMS data, the average delay time for the onset of NW formation was as long as 40 min. This time is defined as the time from the beginning of the deposition to the moment when the desorption flux reduces by one half of the difference between the desorption fluxes during the incubation and elongation stages, respectively. Taking into account this delay time, we find that the axial growth rate during the growth of the NW template was as high as 10.6 nm/min. This value is very close to $\Phi_N$ indicating that, even though the growth was carried out under high N excess ($\Phi_{Ga}$ = 4.8 nm/min and $\Phi_N$ = 10.8 nm/min), the axial growth rate was N-limited. During the second step, $\Phi_{Ga}^{QMS}$ strongly depends on $\Phi_{Ga}$. If $\Phi_{Ga}$ is decreased with respect to the value used for the NW template, $\Phi_{Ga}^{QMS}$ decreases and steady-state growth conditions are re-established immediately. In contrast, an increase in $\Phi_{Ga}$ during the second step results in a higher $\Phi_{Ga}^{QMS}$ and in a transitory behavior with a transient time proportional to $\Phi_{Ga}$. The steady-state values, reached after the transient period of the second stage of the growth (see Figure Figure 1) are shown in Figure Figure 2. It follows from these data that approximately half of the impinging

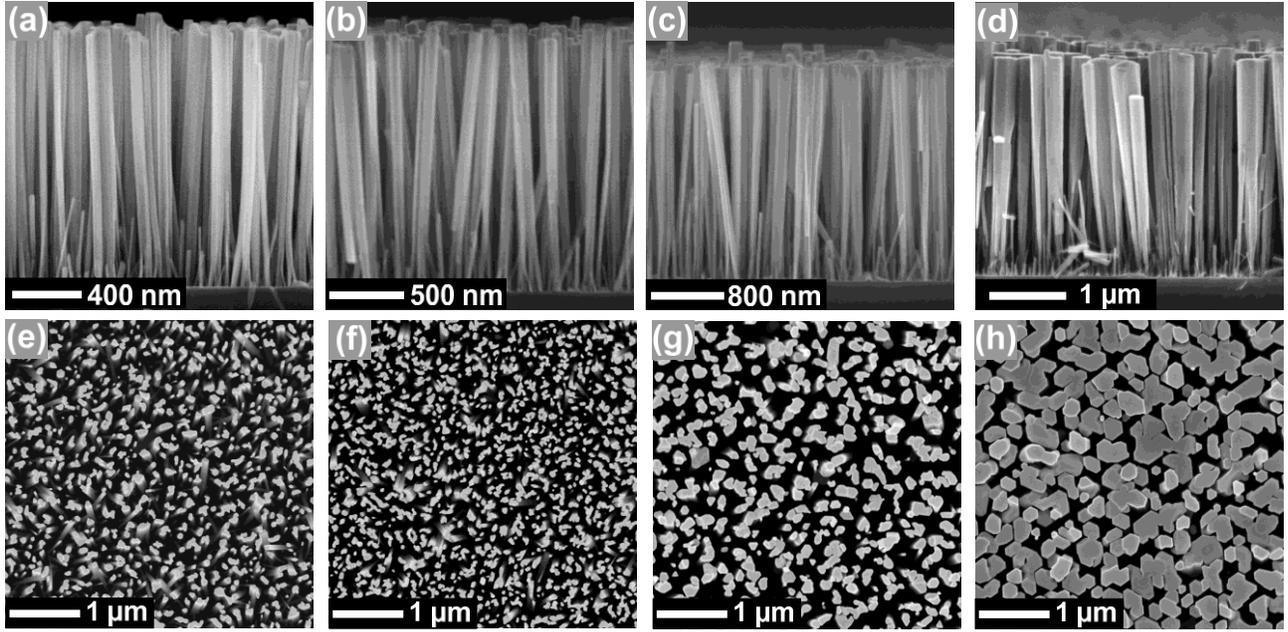

Figure 3: Cross-sectional and plan-view SEM images, respectively, of the GaN NW template grown with $\Phi_{Ga}$ = 4.8 nm/min [(a)/(e)] and the overgrown NW templates using $\Phi_{Ga}$ = 2.1 nm/min [(b)/(f)], $\Phi_{Ga}$ = 12 nm/min [(c)/(g)], and $\Phi_{Ga}$ = 26 nm/min [(d)/(h)]. Note the different scale in Figures (a)–(d).

Ga flux is desorbed.

The morphology of the different samples was investigated by SEM. Figure Figure 3 presents cross-sectional and plan-view SEM images of the NW template and some illustrative samples. One can see in Figures Figure 3(b) and Figure 3(f) that a decrease in $\Phi_{Ga}$ does not significantly influence the morphology of the GaN NWs, which simply become longer during the second step. The situation is entirely different when $\Phi_{Ga}$ is increased. In this case the NWs do not only become longer but also markedly thicker as it can be clearly observed in the plan-view SEM images [Figures Figure 3(g) and Figure 3(h)].

Figure Figure 4 shows that the NW radius does not increase uniformly along the whole NW, but mainly in the upper part of the NWs because the shadowing by the neighboring NWs prevents impinging atoms from reaching the bottom part of the NW sidewalls. The radius of the NWs is seen to continuously increase during the second growth step until reaching a certain equilibrium value. Afterward, the NWs elongate preserving the radius. This variation in NW morphology

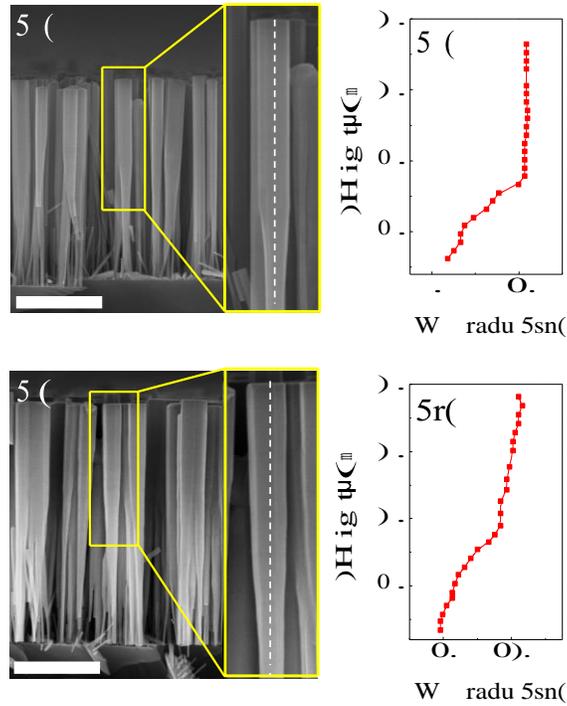

Figure 4: Cross-sectional SEM images of the overgrown NW templates using $\Phi_{Ga}$ = 12 nm/min (a) and $\Phi_{Ga}$ = 26 nm/min (c). The scale bar in (a) and (c) has a length of 1 $\mu$m. (b) and (d) NW radius variation during the second step of the growth. The NW radius was measured along the dashed lines shown in the insets of (a) and (c).

during the second growth step offers a natural explanation for the transitory behavior of $\Phi_{Ga}^{QMS}$ observed by QMS (Figure Figure 1) since Ga adatom incorporation and desorption probabilities are expected to depend on NW shape and size.[38]

The axial growth rate during the second growth step was calculated as the difference between the total average NW height and the nominal height of the NW template divided by the growth time of the second step. Figure Figure 5 shows the variation of the axial growth rate with $\Phi_{Ga}$. The axial growth rate reveals two distinct behaviors depending on whether $\Phi_{Ga}$ is decreased or increased with respect to the value used for the NW template. A decrease in $\Phi_{Ga}$ results in a reduction of the axial growth rate which becomes Ga-limited. Nevertheless, the axial growth rate remains significantly higher than $\Phi_{Ga}$, thus revealing a strong diffusion of Ga adatoms towards the NW tip in agreement with previous studies.[35,39,40,49–53] In contrast, an increase in $\Phi_{Ga}$ does not influence the axial growth rate which remains N-limited. Consequently, N losses caused by GaN decomposition as well as N adatom diffusion along the side surface towards the NW tip are negligible processes during the growth of GaN NWs.

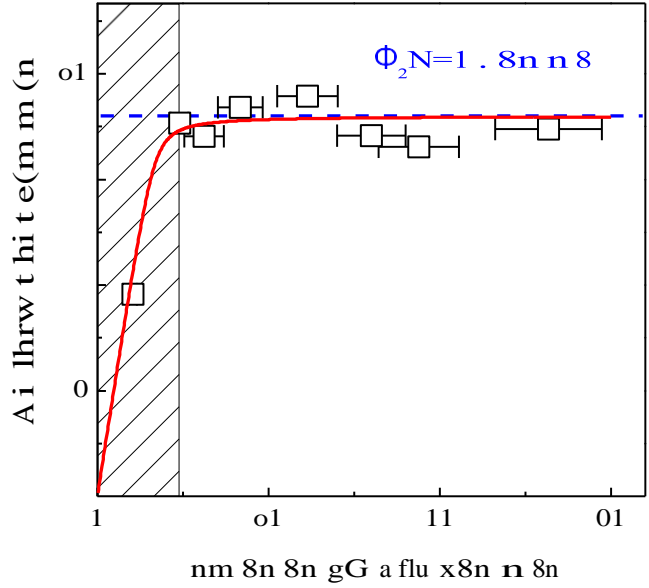

Figure 5: Axial growth rate during the second step of the growth as a function of the impinging Ga flux $\Phi_{Ga}$. The hatched area corresponds to values of $\Phi_{Ga}$ lower than the one used for growing the NW template (4.8 nm/min) and the dashed line indicates the impinging N flux $\Phi_N$. The squares represent the experimental data and the solid line is the model calculation by Eq. Eq. (5). The experimental data for the NW template is also included ($\Phi_{Ga} = 4.8$ nm/min).

In the literature, a Ga-limited growth rate is widely reported,[26,39,50,54,55] but a N-limited growth rate has also been observed by several groups.[33,50–52,56] The present results demonstrate that the axial growth rate of GaN NWs can be limited by either Ga or N depending on the specific growth parameters and geometrical factors, as further discussed below. Our results are in qualitative agreement with those reported by Songmuang *et al.* in Ref. 52, who also observed a transition from Ga- to N-limited growth regimes with increasing $\Phi_{Ga}$ in single GaN NWs investigated via a marker technique.

The top view SEM images [see Figures Figure 3(e)–Figure 3(h)] facilitate an accurate determination of the area fraction $\varphi$ covered by the NWs, as displayed in Figure Figure 6(a). For small Ga fluxes, also the NW density and radius can be determined quite accurately, while the progressive NW coalescence for larger fluxes induces a larger error. Figures Figure 6(b) and Figure 6(c) present the densities of coalesced NWs and their average radii defined as $\sqrt{A/\pi}$ with the NW area $A$. The NW density decreases with increasing Ga fluxes due to the progressive increase of NW coalescence. The error bars in Figure Figure 6(c) are the standard deviations of the NW radius distributions and reflect the decreasing accuracy in the mean NW radius determination with increasing coalescence.

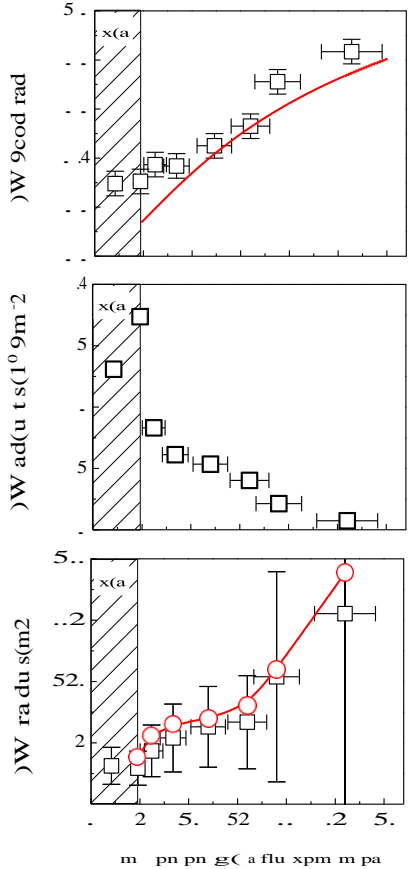

Figure 6: Average parameters of the NW distribution for the second stage of growth: (a) fraction of the surface area covered by NWs, (b) NW density reduction due to coalescence, (c) corresponding NW radius increase. The squares represent experimental data and the solid line and the circles in (a) and (c) are the model calculations by Eq. Eq. (7) and Eq. Eq. (8), respectively. The solid line in (c) is a guide to the eye. The experimental data for the NW template were also included. The hatched area corresponds to values of $\Phi_{Ga}$ lower than the one used for growing the NW template ($\Phi_{Ga} = 4.8$ nm/min).

Inspired by the above observations, we propose a growth model that takes into account both the diffusion of Ga adatoms towards the NW tip and the finite amount of active N available for the growth. The model describes the growth of GaN NWs after the nucleation stage is completed.[42] In accordance with numerous experiments,[35,39,40,49–53] the axial growth is assumed to proceed due to Ga atoms impinging both directly on the top facet of the NW and on the side facets where they may reach the top facet by surface diffusion. In contrast, N adatom diffusion along the NW sidewalls to the tip is considered to be negligible, as follows from the experimental results shown above and those reported in Refs. 50 and 52. In addition, density functional theory calculations show that the N-terminated *M*-plane that constitute the NW sidewall facets is unstable against the formation and desorption of $N_2$ molecules.[57] Hence, axial NW growth must proceed due to N atoms directly impinging on the top facet.

Under steady state growth conditions, we can write the following rate equation for Ga atoms per unit area at the NW top:

$$\Phi_{Ga} + \Phi_{Ga}^{diff} - \Gamma - n_{Ga}/\tau_{Ga} = 0. \tag{1}$$

The first two terms in this equation, $\Phi_{Ga}$ and $\Phi_{Ga}^{diff}$, describe Ga atoms arriving at the NW top facet as a result of the directly impinging Ga flux and the diffusion of Ga adatoms from the NW side facets, respectively. The other two terms, $\Gamma$ and $n_{Ga}/\tau_{Ga}$, describe the consumption and loss of Ga due to its incorporation at the NW top facet and its desorption, respectively. Here $n_{Ga}$ is the Ga adatom density at the top facet and $\tau_{Ga}$ is the Ga adatom lifetime at this facet.

The flux from the side facets $\Phi_{Ga}^{diff}$ is due to Ga atoms impinging on the side facets at distances from the NW top that do not exceed some characteristic distance $L$. Those Ga atoms impinging at larger distances desorb or contribute to NW radial growth.[33,40] For a low-density NW array, $L$ is equal to the diffusion length at the side surface, while for a dense NW array, $L$ will eventually be limited by the shadowing due to neighboring NWs. In any case, the flux from the side facets is proportional to $LR\Phi_{Ga}$. Since this flux is redistributed over the top facet area $\pi R^2$, the flux from the side facets is given by $\Phi_{Ga}^{diff} = \beta (L/R)\Phi_{Ga}$, where $\beta$ is a free parameter that includes all geometric factors. Eq. Eq. (1) can now be written as

$$\frac{n_{Ga}}{\tau_{Ga}} = \left(1 + \beta \frac{L}{R}\right)\Phi_{Ga} - \Gamma. \tag{2}$$

The growth rate is proportional to the densities of both elements, $n_{Ga}$ and $n_N$, at the top NW

facet,

$$\Gamma = K n_{Ga} n_N, \quad (3)$$

where $K$ is the reaction rate constant. The N adatom density at the top facet $n_N$ is described analogously to Eq. Eq. (1), with the flux from the side surfaces being absent due to the reasons previously discussed. Hence,

$$\Phi_N - \Gamma - \frac{n_N}{\tau_N} = 0, \quad (4)$$

where $\tau_N$ is the lifetime of N adatoms at the top facet. Substituting $n_{Ga}$ and $n_N$ from Eq. Eq. (2) and Eq. Eq. (4) into Eq. Eq. (3), one obtain a quadratic equation for $\Gamma$. The solid line in Figure Figure 5 shows the axial growth rate $\Gamma$ obtained from the solution of these equations with $g = (K\tau_{Ga}\tau_N)^{-1} = 0.2$ nm/min and $\beta\, L/R = 2.2$. As can be observed, Eq. Eq. (2)–Eq. Eq. (4) properly describe the transition between the Ga- and N-limited growth regimes. The ratio $L/R$ was taken constant despite the fact that it depends on the Ga flux, as discussed in detail below. However, its variation has no significant impact on the transition between Ga- and N-limited growth regimes.

It is instructive to rewrite Eq. Eq. (2)–Eq. Eq. (4) in the following compact form:

$$g \frac{\Gamma}{\Phi_N - \Gamma} = \left(1 + \beta \frac{L}{R}\right) \Phi_{Ga} - \Gamma. \quad (5)$$

In the limit of negligible Ga desorption and excess of N, i.e., $g \ll \Gamma \ll \Phi_N$, the NW axial growth rate is Ga-limited and Eq. Eq. (5) reduces to

$$\Gamma = \left(1 + \beta \frac{L}{R}\right) \Phi_{Ga}. \quad (6)$$

This equation is a well-known growth law commonly used to describe the diffusion-induced growth of semiconductor NWs.[58–63] It also has been applied to explain the distribution of the aspect ratio of GaN NWs.[40,49] However, we stress that this law is not applicable when the growth rate approaches the N flux ($\Phi_N \approx \Gamma$), as for most of the samples shown in Figure Figure 5 where the axial growth rate is not Ga-, but N-limited.

The present model does not only allow for a quantitative description of the axial growth rate of GaN NWs, but also provides an explanation for the variation of the NW radius with the impinging Ga flux shown in Figure Figure 6(c). Consider first NWs grown with constant Ga supply, such as the template, where the NWs exhibit an essentially constant radius along the growth direction. It then follows from Eq. Eq. (2) that the Ga adatom density $n_{Ga}$ at the NW top must be constant during growth. Second, the results shown in Figure Figure 6(c) reveal that an increase in the Ga flux promotes NW radial growth. Since the side facets of GaN NWs are atomically flat,[15,37] radial growth should proceed layer by layer, with nucleation of a two-dimensional island and further expansion of the island by step-flow. In principle, the islands may nucleate homogeneously at the facets, or at the edges between the side surface and the substrate or the top facet. Homogeneous nucleation at the facets is unlikely because it would cause cone-shaped NWs[64] and also because the NW sidewalls are not simultaneously exposed to Ga and N fluxes.[38] Nucleation at the edge between the side surface and the substrate is also unlikely for dense NW arrays, since this edge is shadowed from the impinging fluxes. Hence, the edge between the top and the side facets is the most plausible nucleation site for radial growth. Two-dimensional island nucleation requires that the adatom density exceeds a certain critical value. Since the Ga adatom density at the top of the side facet $n_f$ is related to the Ga adatom density at the top facet $n_{Ga}$ by the equality of the chemical potentials, $\mu_f(n_f) = \mu_{Ga}(n_{Ga})$, we conclude that the increase in NW radius with increasing Ga fluxes shown in Figure Figure 6(c) occurs because the Ga adatom concentration at the top facet exceeds a certain threshold when the Ga flux is increased.

During subsequent radial growth, the Ga flux from the side surface to the top facet $\Phi_{Ga}^{diff}$ continuously decreases due to the increase of $R$. Eq. Eq. (2) predicts that this reduction of the effective flux from the side facets reduces $n_{Ga}$ until it eventually becomes lower than the critical value for island nucleation at the NW side facets. At that point, radial growth ceases and the NW elongates preserving its radius, as evidenced by Figure Figure 4. Eq. Eq. (2) with $\Gamma \approx \Phi_N$ and the critical value of $n_{Ga}$ hence define the equilibrium radius $R$ for a given Ga flux.

For a further quantitative comparison of the experimental data with our theoretical model, we assume that the characteristic distance $L$ is limited by shadowing from neighboring NWs. In this case, $L$ decreases as the fraction $\varphi$ of the surface covered by NWs increases due to the increase in NW radius. In our experiment, the height of the side surface $L$ exposed to the impinging molecular beams is approximately equal to the distance between NW surfaces, $L \approx l - 2R$, where $l$ is the distance between the centers of NWs. Since the ratio $\pi R^2/l^2$ equals the coverage $\varphi$, we can write $L/R = \sqrt{\pi/\varphi} - 2$. Substituting this ratio in Eq. Eq. (2) and assuming that the desorbing Ga flux from the NW top facet under steady state growth conditions $\Phi_{Ga}^{des} = n_{Ga}/\tau_{Ga}$ is a constant that does not depend on the Ga flux, we can express the area fraction covered by the NWs as

$$\varphi = \pi \left[ \frac{1}{\beta} \left( \frac{\Phi_{Ga}^{des} + \Gamma}{\Phi_{Ga}} - 1 \right) + 2 \right]^{-2}. \qquad (7)$$

Figure Figure 6(a) compares the experimental dependence of the coverage on the Ga flux with the one calculated by Eq. Eq. (7) with $\Phi_{Ga}^{des} = 18$ nm/min and $\beta = 1.8$ for Ga fluxes higher than the one used for growing the NW template. Since $\Phi_{Ga}^{des}$ is the Ga flux desorbed per unit area of the NW top facet, the total desorbed Ga flux is approximately $\Phi_{Ga}^{QMS} = \varphi \Phi_{Ga}^{des}$. This quantity, calculated by Eq. Eq. (7) with the same value $\Phi_{Ga}^{des} = 18$ nm/min, is compared in Figure Figure 2 with the actual desorbing Ga flux measured by QMS.

Within the approximation $L = l - 2R$, the NW radius $R$ can be expressed using Eq. Eq. (2) through the distance between NW centers $l$ as

$$R = l \left[ \frac{1}{\beta} \left( \frac{\Phi_{Ga}^{des} + \Gamma}{\Phi_{Ga}} - 1 \right) + 2 \right]^{-1}. \qquad (8)$$

Figure Figure 6(c) shows the NW radius calculated by Eq. Eq. (8) with the same parameters as above, $\Phi_{Ga}^{des} = 18$ nm/min and $\beta = 1.8$, and taking the values for $\Gamma$ shown in Figure Figure 5. The distance between NW centers is calculated as $l = \rho^{-1/2}$ where $\rho$ is the measured NW density shown in Figure Figure 6(b).

Apart from the quite satisfactory quantitative description of the axial and radial growth of GaN NWs shown above, our model also allows us to explain the necessity of effectively N-rich growth conditions for the growth of GaN NWs. In order to obtain well separated NWs, it is required that $R < l/2$. Taking into account that $\Gamma = \Phi_N$ during radial growth, Eq. Eq. (8) then results in the inequality $\Phi_N > \Phi_{Ga} - \Phi_{Ga}^{des}$ as requirement for NW growth. In other words, the impinging N flux $\Phi_N$ must exceed the effective Ga flux for obtaining GaN NWs.

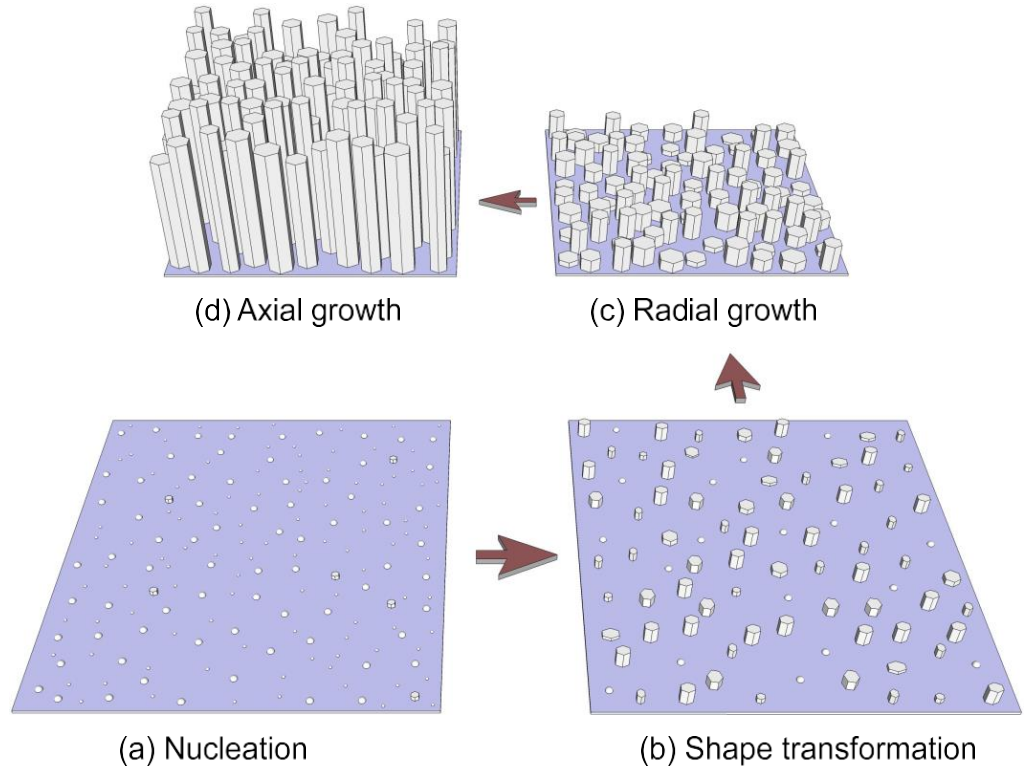

Figure 7: Schematic illustration of the different stages during the spontaneous nucleation and growth of GaN NWs on Si(111). (a) Nucleation stage with the formation of stable spherical cap-shaped 3D islands. (b) Shape transition to the NW-like morphology. (c) NW elongation and radial growth until the Ga adatom density at the NW top facet is reduced below the threshold for two-dimensional island nucleation at the NW side facets. (d) Final stage where GaN NWs elongate preserving their radius.

Figure Figure 7 displays the evolution of GaN NWs on Si(111) in PA-MBE from the nucleation stage to their final configuration as a dense NW array. The sketch results from a combination of the nucleation model published by Consonni *et al.* in Ref. 42 with the kinetic growth model introduced here. The growth process begins with the incubation stage, where unstable 3D GaN nuclei are spontaneously forming and disintegrating on the surface. Stable spherical cap-shaped

3D islands grow until reaching a certain critical radius ($\sim$ 5 nm). After this radius is reached, the shape transition to the NW-like morphology with a flat $(000\bar{1})$ top facet and $(10\bar{1}0)$ side facets occurs. At this stage, each NW is able to collect a significant amount of Ga from the substrate since shadowing is not yet operative. The Ga adatom concentration on the NW top facet will thus exceed the threshold value for two-dimensional island formation at the NW side facets, and radial growth will set in. At this stage, further nucleation at the edge between the substrate and the side facets is also possible because that edge is not shadowed yet.[41] As time evolves, the NW radius increases, the NWs elongate and shadowing progressively sets in. As the NW radius increases, the Ga adatom density at the top facet continuously decreases until its value is below the threshold for two-dimensional island formation at the side facets. At that point the radial growth ceases and the NWs elongate preserving their radius if the Ga atoms impinging on the side facets can reach the NW top facet by surface diffusion. Otherwise, the NW radius can increase during growth but with a growth rate that is more than one order of magnitude lower than the axial growth rate, as reported in Refs. 33, 40 and 41. While the nucleation stage strongly depends on the substrate,[42,65] the kinetic growth model presented here is of entirely general nature.

To summarize and conclude, we have shown that despite the effectively N-rich conditions during growth of GaN NWs, the finite amount of active N available for the growth may limit their axial growth rate. This effect is caused by the different diffusion lengths of Ga and N adatoms along the *M*-plane facets that constitute the NW sidewalls. Whereas Ga adatoms are highly mobile and able to reach the NW tip, N adatoms are unstable against the formation and desorption of $N_2$ molecules. Consequently, as a result of Ga diffusion on the side facets, the amount of Ga reaching the NW tip can exceed the amount of active N available for the growth even for impinging III/V flux ratios lower than 1. Under such conditions, the excess Ga does not accumulate at the NW tip but induces radial growth of the NWs. This process is a self-regulated one, since the amount of Ga available for the growth at the NW tip decreases with increasing NW radius. The NW radial growth ceases when the Ga adatom density at the NW top is reduced to a certain critical value. If this condition cannot be satisfied, a compact layer forms due to NW coalescence. Therefore, our kinetic growth model provides a natural explanation for two well known but so far not fully understood experimental results: (i) the necessity of using effectively N-rich growth conditions for the spontaneous growth of GaN NWs, and (ii) the increase in NW radius with increasing III/V flux ratio. In addition, we have also shown that the axial growth rate of GaN NWs can be either increased or decreased during NW growth by changing the impinging Ga flux, but the NW radius can only be increased. The present work thus does not only provide a comprehensive description of the spontaneous formation of GaN NWs in PA-MBE, but also represents a useful guide for modifying *in-situ* the morphological properties of group-III nitride NW heterostructures, which is of great interest for the fabrication of NW-based devices. Furthermore, within a more general framework, the results presented here demonstrate that in order to fully understand the diffusion-induced growth of compound semiconductor NWs, it is necessary to consider not only the element limiting the growth and reaching the NW tip by surface diffusion but also the actual finite amount of the element that is present in abundance.

# Acknowledgement

The authors would like to thank Sergey Sadofev for his help to calibrate the MBE system, Anne-Kathrin Bluhm for the SEM images, Hans-Peter Schönherr for the maintenance of the MBE system and Caroline Chèze for a critical reading of the manuscript. Karl K. Sabelfeld has been supported by Deutsche Forschungsgemeinschaft (DFG) grant KA 3262/2–2, and Enrique Calleja would like to acknowledge partial financial support by the EU-FP7 Contract GECCO 280694-2 and by the Spanish project CAM/P2009/ESP-1503.

# References

(1) Wagner, R. S.; Ellis, W. C. *Appl. Phys. Lett.* **1964**, *4*, 89–90.

(2) Lauhon, L. J.; Gudiksen, M. S.; Lieber, C. M. *Philosophical transactions. Series A, Mathematical, physical, and engineering sciences* **2004**, *362*, 1247–60.

(3) Perea, D. E.; Allen, J. E.; May, S. J.; Wessels, B. W.; Seidman, D. N.; Lauhon, L. J. *Nano Lett.* **2006**, *6*, 181–5.

(4) Glas, F.; Harmand, J.-C.; Patriarche, G. *Phys. Rev. Lett.* **2007**, *99*, 146101.

(5) Krogstrup, P.; Curiotto, S.; Johnson, E.; Aagesen, M.; Nygård, J.; Chatain, D. *Phys. Rev. Lett.* **2011**, *106*, 125505.

(6) Yu, X.; Wang, H.; Lu, J.; Zhao, J.; Misuraca, J.; Xiong, P.; von Molnár, S. *Nano Lett.* **2012**, *12*, 5436–42.

(7) Calleja, E.; Ristić, J.; Fernández-Garrido, S.; Cerutti, L.; Sánchez-García, M. A.; Grandal, J.; Trampert, A.; Jahn, U.; Sánchez, G.; Griol, A.; Sánchez, B. *Phys. Status Solidi B* **2007**, *244*, 2816–2837.

(8) Bertness, K. A.; Member, S.; Sanford, N. A.; Davydov, A. V. *IEEE J. Sel. Topics in Quantum Electron.* **2011**, *17*, 847–858.

(9) Schmidt-Mende, L.; Macmanus-Driscoll, J. L. *Materials Today* **2007**, *10*, 40–48.

(10) Perillat-Merceroz, G.; Thierry, R.; Jouneau, P.-H.; Ferret, P.; Feuillet, G. *Nanotechnology* **2012**, *23*, 125702.

(11) Fernández-Garrido, S.; Kong, X.; Gotschke, T.; Calarco, R.; Geelhaar, L.; Trampert, A.; Brandt, O. *Nano Lett.* **2012**, *12*, 6119.

(12) Calleja, E.; Sánchez-García, M.; Sánchez, F.; Calle, F. B.; Naranjo, F.; Muñoz, E.; Jahn, U.; Ploog, K. *Phys. Rev. B* **2000**, *62*, 16826–16834.

(13) Lee, D. J.; Park, J. Y.; Yun, Y. S.; Hong, Y. S.; Moon, J. H.; Lee, B.-T.; Kim, S. S. *J. Cryst. Growth* **2005**, *276*, 458–464.

(14) Cerutti, L.; Ristić, J.; Fernández-Garrido, S.; Calleja, E.; Trampert, A.; Ploog, K. H.; Lazic, S.; Calleja, J. M. *Appl. Phys. Lett.* **2006**, *88*, 213114.

(15) Stoica, T.; Sutter, E.; Meijers, R. J.; Debnath, R. K.; Calarco, R.; Lüth, H.; Grützmacher, D. *Small* **2008**, *4*, 751–4.

(16) Brandt, O.; Pfüller, C.; Chèze, C.; Geelhaar, L.; Riechert, H. *Phys. Rev. B* **2010**, *81*, 45302.

(17) Lin, Y.; Chen, W.-J.; Lu, J. Y.; Chang, Y. H.; Liang, C.-T.; Chen, Y. F.; Lu, J.-Y. *Nanoscale research letters* **2012**, *7*, 401.

(18) Kikuchi, A.; Kawai, M.; Tada, M.; Kishino, K. *Jpn. J. Appl. Phys.* **2004**, *43*, L1524–L1526.

(19) Aluri, G. S.; Motayed, A.; Davydov, A. V.; Oleshko, V. P.; Bertness, K. A.; Sanford, N. A.; Rao, M. V. *Nanotechnology* **2011**, *22*, 295503.

(20) Consonni, V.; Rey, G.; Bonaimé, J.; Karst, N.; Doisneau, B.; Roussel, H.; Renet, S.; Bellet, D. *Appl. Phys. Lett.* **2011**, *98*, 111906.

(21) Wang, D.; Pierre, A.; Kibria, M. G.; Cui, K.; Han, X.; Bevan, K. H.; Guo, H.; Paradis, S.; Hakima, A. R.; Mi, Z. *Nano Lett.* **2011**, *11*, 2353.

(22) Li, S.; Waag, A. *J. Appl. Phys.* **2012**, *111*, 071101.

(23) Lu, Y.-J.; Kim, J.; Chen, H.-Y.; Wu, C.; Dabidian, N.; Sanders, C. E.; Wang, C.-Y.; Lu, M.-Y.; Li, B.-H.; Qiu, X.; Chang, W.-H.; Chen, L.-J.; Shvets, G.; Shih, C.-K.; Gwo, S. *Science* **2012**, *337*, 450–453.

(24) Wallys, J.; Teubert, J.; Furtmayr, F.; Hofmann, D. M.; Eickhoff, M. *Nano Lett.* **2012**, *12*, 6180.

(25) Koblmüller, G.; Fernández-Garrido, S.; Calleja, E.; Speck, J. S. *Appl. Phys. Lett.* **2007**, *91*, 161904.

(26) Fernández-Garrido, S.; Grandal, J.; Calleja, E.; Sánchez-García, M. A.; López-Romero, D. *J. Appl. Phys.* **2009**, *106*, 126102.

(27) Kong, X.; Ristić, J.; Sánchez-García, M. A.; Calleja, E.; Trampert, A. *Nanotechnology* **2011**, *22*, 415701.

(28) Hestroffer, K.; Bougerol, C.; Leclere, C.; Renevier, H.; Daudin, B. *Phys. Rev. B* **2011**, *84*, 245302.

(29) den Hertog, M. I.; González-Posada, F.; Songmuang, R.; Rouviere, J. L.; Fournier, T.; Fernandez, B.; Monroy, E. *Nano Lett.* **2012**, *12*, 5691–6.

(30) de la Mata, M.; Magen, C.; Gazquez, J.; Utama, M. I. B.; Heiss, M.; Lopatin, S.; Furtmayr, F.; Fernández-Rojas, C. J.; Peng, B.; Morante, J. R.; Rurali, R.; Eickhoff, M.; Fontcuberta i Morral, A.; Xiong, Q.; Arbiol, J. *Nano Lett.* **2012**, *12*, 2579–86.

(31) Schuster, F.; Furtmayr, F.; Zamani, R.; Magén, C.; Morante, J. R.; Arbiol, J.; Garrido, J. A.; Stutzmann, M. *Nano Lett.* **2012**, *12*, 2199–204.

(32) Park, Y.; Lee, S.; Oh, J.; Park, C.; Kang, T. *J. Cryst. Growth* **2005**, *282*, 313–319.

(33) Tchernycheva, M.; Sartel, C.; Cirlin, G. E.; Travers, L.; Patriarche, G.; Harmand, J.-C.; Dang, L. S.; Renard, J.; Gayral, B.; Nevou, L.; Julien, F. *Nanotechnology* **2007**, *18*, 385306.

(34) Consonni, V.; Trampert, A.; Geelhaar, L.; Riechert, H. *Appl. Phys. Lett.* **2011**, *99*, 033102.

(35) Calarco, R.; Meijers, R. J.; Debnath, R. K.; Stoica, T.; Sutter, E.; Lüth, H. *Nano Lett.* **2007**, *7*, 2248–51.

(36) Bertness, K. A.; Roshko, A.; Mansfield, L. M.; Harvey, T. E.; Sanford, N. A. *J. Cryst. Growth* **2008**, *310*, 3154–3158.

(37) Ristić, J.; Calleja, E.; Fernández-Garrido, S.; Cerutti, L.; Trampert, A.; Jahn, U.; Ploog, K. H. *J. Cryst. Growth* **2008**, *310*, 4035–4045.

(38) Foxon, C. T.; Novikov, S.; Hall, J.; Campion, R.; Cherns, D.; Griffiths, I.; Khongphetsak, S. *J. Cryst. Growth* **2009**, *311*, 3423–3427.

(39) Consonni, V.; Dubrovskii, V. G.; Trampert, A.; Geelhaar, L.; Riechert, H. *Phys. Rev. B* **2012**, *85*, 155313.

(40) Dubrovskii, V. G.; Consonni, V.; Geelhaar, L.; Trampert, A.; Riechert, H. *Appl. Phys. Lett.* **2012**, *100*, 153101.

(41) Dubrovskii, V.; Consonni, V.; Trampert, A.; Geelhaar, L.; Riechert, H. *Phys. Rev. B* **2012**, *85*, 165317.

(42) Consonni, V.; Hanke, M.; Knelangen, M.; Geelhaar, L.; Trampert, A.; Riechert, H. *Phys. Rev. B* **2011**, *83*, 035310.

(43) Heying, B.; Averbeck, R.; Chen, L. F.; Haus, E.; Riechert, H.; Speck, J. S. *J. Appl. Phys.* **2000**, *88*, 1855.

(44) Koblmüller, G.; Averbeck, R.; Riechert, H.; Pongratz, P. *Phys. Rev. B* **2004**, *69*, 035325.

(45) Brown, J. S.; Koblmüller, G.; Wu, F.; Averbeck, R.; Riechert, H.; Speck, J. S. *J. Appl. Phys.* **2006**, *99*, 074902.

(46) Abramoff, M. D.; Magelhaes, P. J.; Ram, S. J. *Biophotonics International* **2004**, *11*, 36.

(47) Chèze, C.; Geelhaar, L.; Trampert, A.; Riechert, H. *Appl. Phys. Lett.* **2010**, *97*, 043101.

(48) Limbach, F.; Caterino, R.; Gotschke, T.; Stoica, T.; Calarco, R.; Geelhaar, L.; Riechert, H. *AIP Advances* **2012**, *2*, 012157.

(49) Debnath, R. K.; Meijers, R. J.; Richter, T.; Stoica, T.; Calarco, R.; Lüth, H. *Appl. Phys. Lett.* **2007**, *90*, 123117.

(50) Songmuang, R.; Landré, O.; Daudin, B. *Appl. Phys. Lett.* **2007**, *91*, 251902.

(51) Landré, O.; Songmuang, R.; Renard, J.; Bellet-Amalric, E.; Renevier, H.; Daudin, B. *Appl. Phys. Lett.* **2008**, *93*, 183109.

(52) Songmuang, R.; Ben, T.; Daudin, B.; González, D.; Monroy, E. *Nanotechnology* **2010**, *21*, 295605.

(53) Galopin, E.; Largeau, L.; Patriarche, G.; Travers, L.; Glas, F.; Harmand, J.-C. *Nanotechnology* **2011**, *22*, 245606.

(54) Sánchez-García, M.; Calleja, E.; Monroy, E.; Sánchez, F.; Calle, F.; Muñoz, E.; Beresford, R. *J. Cryst. Growth* **1998**, *183*, 23–30.

(55) Chèze, C.; Geelhaar, L.; Jenichen, B.; Riechert, H. *Appl. Phys. Lett.* **2010**, *97*, 153105.

(56) Yoshizawa, M.; Kikuchi, A.; Mori, M.; Fujita, N.; Kishino, K. *Jpn. J. Appl. Phys.* **1997**, *36*, L459–L462.

(57) Lymperakis, L.; Neugebauer, J. *Phys. Rev. B* **2009**, *79*, 241308.

(58) Ruth, V.; Hirth, J. P. *J. Chem. Phys.* **1964**, *41*, 3139.

(59) Givargizov, E. I. *J. Cryst. Growth* **1975**, *31*, 20–30.

(60) Schubert, L.; Werner, P.; Zakharov, N. D.; Gerth, G.; Kolb, F. M.; Long, L.; Gösele, U.; Tan, T. Y. *Appl. Phys. Lett.* **2004**, *84*, 4968.

(61) Johansson, J.; Svensson, C. P. T.; Mårtensson, T.; Samuelson, L.; Seifert, W. *J. Phys. Chem. B* **2005**, *109*, 13567–13571.

(62) Dubrovskii, V.; Sibirev, N.; Suris, R.; Cirlin, G.; Ustinov, V.; Tchernysheva, M.; Harmand, J. *Semiconductors* **2006**, *40*, 1075–1082.

(63) Plante, M.; LaPierre, R. *J. Cryst. Growth* **2006**, *286*, 394 – 399.

(64) Dubrovskii, V. G.; Sibirev, N. V.; Cirlin, G. E.; Tchernycheva, M.; Harmand, J. C.; Ustinov, V. M. *Phys. Rev. E* **2008**, *77*, 031606.

(65) Consonni, V.; Knelangen, M.; Geelhaar, L.; Trampert, A.; Riechert, H. *Phys. Rev. B* **2010**, *81*, 085310.

# Table of Contents Graphic

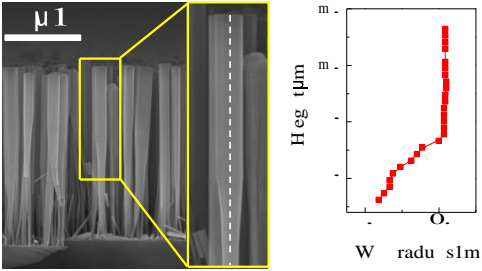